\documentclass[preprint,superscriptaddress,preprintnumbers,showpacs]{revtex4}
\usepackage{amssymb}
\usepackage{graphicx}
\usepackage{dcolumn}
\usepackage{bm}
\usepackage{slashed,color}

\begin{document}

\newcommand*{\sjtu}{INPAC, SKLPPC and Department of Physics, Shanghai Jiao Tong University,  Shanghai, China}\affiliation{\sjtu}
\newcommand*{\NTU}{CTS, CASTS and Department of Physics, \\National Taiwan University, Taipei, Taiwan}\affiliation{\NTU}
\newcommand*{\ncts}{National Center for Theoretical Sciences, Hsinchu, Taiwan}\affiliation{\ncts}
\newcommand*{\kias}{School of Physics, Korea Institute of Advanced Study, Seoul, Korea}\affiliation{\kias}

\hfill{}

\title{A $\gamma\gamma$ Collider for the 750 GeV Resonant State}
\affiliation{\NTU}
\author{Min He}\email{hemin_sjtu@163.com}\affiliation{\sjtu}
\author{Xiao-Gang He}\email{hexg@phys.ntu.edu.tw}\affiliation{\sjtu}\affiliation{\NTU}\affiliation{\ncts}
\author{Yong Tang}\email{ytang@kias.re.kr}\affiliation{\kias}

\begin{abstract}
Recent data collected by ATLAS and CMS at 13 TeV collision energy of the LHC indicate the existence of a new resonant state $\phi$ with a mass of 750 GeV decaying into two photons $\gamma\gamma$.  The properties of $\phi$ should be studied further at the LHC and also future colliders. Since only $\phi \to \gamma\gamma$ decay channel has been measured, one of the best ways to extract more information about $\phi$ is to use a $\gamma\gamma$ collider to produce $\phi$ at the resonant energy. In this work we show how a $\gamma\gamma$ collider helps to verify the existence of $\phi$ and to provide some of the most important information about the properties of $\phi$, such as branching fractions of $\phi\to V_1V_2$. Here $V_i$ can be $\gamma$, $Z$, or $W^\pm$. We also show that by studying angular distributions of the final $\gamma$'s in $\gamma\gamma \to \phi \to \gamma\gamma$, one can obtain crucial information about whether this state is a spin-0 or a spin-2 state.
\end{abstract}


\date{\today $\vphantom{\bigg|_{\bigg|}^|}$}

\maketitle

\noindent
{\bf Introduction}

Recent data collected by ATLAS and CMS at $\sqrt{s} =13$ TeV collision energy of the LHC indicate the existence of a new resonant state $\phi$ with a mass of 750 GeV decaying into two photons~\cite{exp}. 
The production cross section $\sigma(pp \to \phi \to \gamma\gamma)$ is about 6 fb with the ATLAS data hinting that $\phi$ has a broad width about 45 GeV with a local significance of 3.9$\sigma$, while CMS data favor a narrow width of order 100 MeV with a local significance of 2.6$\sigma$~\cite{exp}.  Combining ATLAS and CMS data, one obtains: $\sigma(pp\to \phi \to \gamma\gamma) = (6\pm 2) $ fb at 13 TeV. With the assumption that $\phi$ is mainly produced by gluon fusion, $gg \to \phi$, one obtains $\Gamma_{gg}\Gamma_{\gamma\gamma}/\Gamma_{total} \sim 1$ MeV. Here $\Gamma_{ii}$ is the partial decay width of $\phi \to ii$. If one also assumes that  $\phi$ dominantly decays into two gluons and is produced by gluon fusion, $gg \to \phi$, one can extract a lower bound of about 1 MeV for $\Gamma_{ \gamma\gamma}$. If one takes the total width to be 45 GeV indicated by ATLAS, the branching ratio of $\phi \to \gamma \gamma$ is only about a few times $10^{-5}$. More data are needed to confirm the existence of this new state. The LHC will continue to run and will soon have more to tell about the properties of $\phi$.  At future colliders, more aspects of this resonant state can be studied. At present, only limited information about $\phi$ is available, namely $\phi$ is produced at $pp$ collision and it decays into $\gamma \gamma$ final state. Regarding $\gamma\gamma$ decay from a state, the situation is similar to that of the 125 GeV Higgs boson discovered at the LHC. Therefore some of the strategies at future collider for the study of $\phi$ can be employed except that the energy has to be increased. In particular we note that a $\gamma\gamma$ collider may be an ideal place to study some of the most important properties of the possible new resonant state $\phi$ similar to the study of Higgs boson properties~\cite{gammagammacollider}. 

The possibility of the existence of $\phi$ has generated a lot of theoretical speculations. 
If $\phi$ exists, it must come from beyond the standard model (SM). We will concentrate on how a $\gamma \gamma$ collider can provide information about 
the properties of the 750 GeV resonant state. The resonant state $\phi$ can be produced through $\gamma\gamma$ collision. The problem of course is that whether it has a large luminosity to generate enough events to study the properties of $\phi$. We confirm previous studies~\cite{theories3,theories4} that 
a $\gamma\gamma$ collider constructed by using the laser backscattering
technique on the electron and positron beams in an $e^+e^-$ collider with center of mass (CM) frame energy of around 1 TeV and integrated luminosity of order one thousand fb$^{-1}$, many properties of $\phi$ can be studied. In particular some of the expected decay modes $\phi \to V_1V_2$ with $V_i = \gamma,\; Z, \;W^\pm$ can be studied cleanly. 

Since $\phi$ can decay into two on-shell photons, according to Landau-Yang theorem, the state cannot be a spin-1 state~\cite{landau}. The two likely possibilities with low spins are spin-0 and spin-2. Further studies are needed to know the spin property of $\phi$~\cite{spin-2, photo-production}. We show that by studying angular distribution of the final $\gamma \gamma$ through on-shell production of $\phi$ and its subsequent decays into a $\gamma \gamma$ pair, one can easily determine whether $\phi$ is a spin-0 or spin-2 state. The work presented closely follow previous study of Higgs boson at a $\gamma\gamma$ collider by one of us (He)~\cite{gammagammacollider}.
\\

\noindent
{\bf  $\gamma\gamma \to \phi$ and $\phi \to X,\; \gamma\gamma$, and $V_1 V_2$}

Assuming that $\phi$ is a spin-0 scalar state, the $\phi \to \gamma\gamma$ decay amplitude $M(J=0, \gamma\gamma)$  has the form: $M(0,\gamma\gamma) ={A}(k_{2\mu}k_{1\nu}-g_{\mu\nu}k_{1}\cdot k_{2}){\varepsilon^{\mu*}(k_{1})}{\varepsilon^{\nu*}(k_{2})}$, 
which gives the decay width 
\begin{eqnarray}
\Gamma_{0,\gamma\gamma} ={{A^{2}}m^3_\phi/{64\pi}}\;.\label{decay}
\end{eqnarray}

If $\phi$ is a spin-0 pseudoscalar state, the decay amplitude has the form: $M(0, \gamma\gamma) = A \epsilon_{\mu\nu\alpha\beta}k_1^\mu \epsilon_1^\nu k_2^\alpha \epsilon^\beta_2$. The expression for $\Gamma_{0,\gamma\gamma}$ has the same form as in eq. (\ref{decay}).

The cross section $\sigma(s)_{0,X}$ for producing an on-shell $\phi$ at a monochromatic $\gamma\gamma$ collider followed by $\phi$ decays into a final state $X$, $\gamma\gamma \to \phi \to X$ with a center of mass (CM) frame energy $\sqrt{s}$ is directly related to the decay width $\Gamma_{0, \gamma \gamma}$~~\cite{Gunion:1992ce}. For $\phi$ be a scalar (also for a pseudoscalar), we have 
\begin{eqnarray}
\sigma(s)_{0, X} = \Gamma_{0, \gamma\gamma} {{8\pi^{2}}\over{m_{\phi}}}\delta(s-m^2_\phi)Br_{0,X} \;,
\end{eqnarray}
where $Br_{0,X}$ is the branching ratio of $\phi$ decays into the final state  $X$.

A $\gamma\gamma$ collider can be constructed by  using the laser backscattering
technique on the electron and positron beams in an $e^+e^-$ collider. For example the $e^+e^-$ ILC collider. Such a collider has been shown to be useful to study beyond SM physics~~\cite{gammagammacollider,He:2006yy}. In this case the energy $E_\gamma$ of the photons are not monochromatic, but have a distribution $f(x=E_\gamma/E_e)$ for a given electron/positron energy $E_e$~~\cite{I. Ginzburg}.
In the $e^+ e^-$ CM frame, the cross section $\sigma^L_{0,X}$ for $\gamma(x_1)\gamma(x_2) \rightarrow \phi \rightarrow X$ is given by
\begin{eqnarray}
\sigma(s)^L_{0,X} &=&\int_{x_{min}}^{x_{max}}dx_1\int_{x_{min}}^{x_{max}}dx_2 \sigma(x_1x_2s)_{0,X} f(x_1)f(x_2)[1+\lambda(x_1)\lambda(x_2)]\;.
\end{eqnarray}
where $x_1$ and $x_2$ are the fractions of photon energy come from $e^-$ and $e^+$ beams. $x_{max} = \xi/(1+\xi)$ with $\xi = 2(1+\sqrt{2})$, and $x_{min} = y/x_{max}$ with $y=m_{\phi}^2/s$.
$\lambda(x)$ is the mean helicity of the $\gamma$ beam which depends on the $e^-$($e^+$) polarization $\lambda_{e^-}$($\lambda_{e^+}$) and the laser polarization $\lambda_l$, and is given by
\begin{eqnarray}
\lambda(x)=\frac{2\pi\alpha^2}{\sigma_c\xi m_e^2 f(x)}\{\lambda_l(1-2r)(1-x+\frac{1}{1-x})+\lambda_{e}\xi r[1+(1-x)(1-2r)^2]\},
\end{eqnarray}
where $r={x}/{\xi(1-x)}$. 
The distribution function $f(x)$ is given by
\begin{eqnarray}
f(x)=\frac{2\pi\alpha^2}{\sigma_c\xi m_e^2}[\frac{1}{1-x}+1-x-4r(1-r)-\lambda_l\lambda_e\xi r(2r-1)(2-x)],
\end{eqnarray}
with
$\sigma_c=\sigma_c^{np}+\lambda_l\lambda_e\sigma_1$ and
\begin{eqnarray}
\sigma_c^{np}&=&\frac{2\pi\alpha^2}{\xi m_e^2}[(1-\frac{4}{\xi}-\frac{8}{\xi^2})
\ln(1+\xi)+\frac{1}{2}+\frac{8}{\xi}-\frac{1}{2(1+\xi)^2}]\;,\nonumber\\
\sigma_1&=&\frac{2\pi\alpha^2}{\xi m_e^2}[(1+\frac{2}{\xi})\ln(1+\xi)-\frac{5}{2}+\frac{1}{1+\xi}-\frac{1}{2(1+\xi)^2}].
\end{eqnarray}
For unpolarized laser, and unpolarized electron/positron beams, $\lambda_l = \lambda_e =0$, $\lambda(x)$ is also zero.

Integrating out $x_2$, we have
\begin{eqnarray}
\sigma(s)^L_{0,X} &=& I(y)\frac{8\pi^2}{m_{\phi}^3}\Gamma_{0,\gamma\gamma}Br_{0,X}\;,
\end{eqnarray}
where the function $I(y)$ is given by
\begin{eqnarray}
I(y)=\int_{x_{min}}^{x_{max}}dx\frac{y}{x}f(x)f(y/x)[1+\lambda(x)\lambda(y/x)].
\end{eqnarray}

\begin{figure}
\includegraphics[width=7cm]{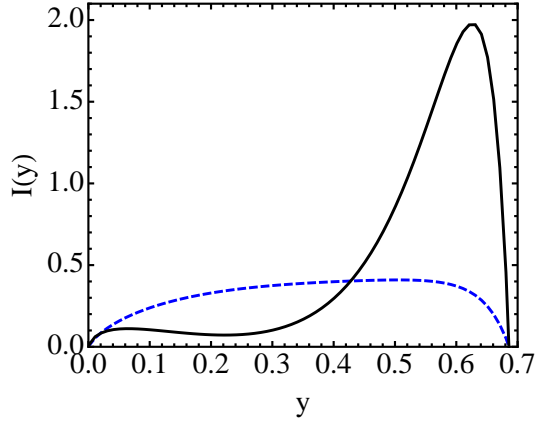}
\caption{$I(y)$ as a function of $y$. Dashed and solid curves are for unpolarized and polarized cases, respectively.}\label{fig1}
\end{figure}

The function $I(y)$ plays a crucial role in gauging at what energy the production of $\phi$ will be maximized.
We plot $I(y)$ for both polarized and unpolarized photon beams, in Figure \ref{fig1}. 
We see that for unpolarized case, $I(y)$ peaks at about 0.4 when $y\simeq 0.6$.
For the polarized case the peak value of $I(y)$ can be enhanced by suitable choice of the 
laser and electron and positron polarizations. For example with $\lambda_e=1$ and $\lambda_l=-1$, the peak value of $I(y)$ can reach 1.8 for $y\sim 0.6$. The enhanced peak value is at the cost of polarization of the beams. On the other hand, unpolarized case may be easier to achieve in practice. 

\begin{figure}
\includegraphics[width=7.5cm]{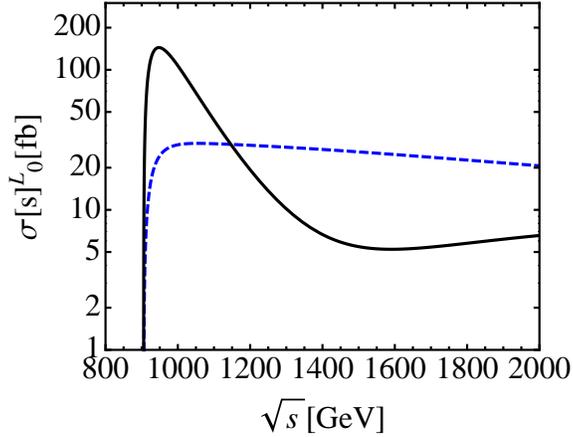}
\caption{The $\gamma\gamma \to \phi$ cross section $\sigma(s)^L_0$ (in unit of fb) as a function of $\sqrt{s}$ with $\Gamma_{0,\gamma\gamma} = 1$MeV. The dashed curve is for unpolarized photon beam and the solid curve is for polarized photon beam.}\label{fig2}
\end{figure}

In Figure \ref{fig2}, we plot the cross section $\sigma^L_0$ for $\gamma\gamma\to \phi$ at a $e^+e^-$ linear collider as a function of $\sqrt{s}$ using $\Gamma_{0, \gamma\gamma} = 1$ MeV. We confirm the results obtained in Ref.~\cite{theories3}. We see that with $\sqrt{s} = 1$ TeV, we see that an integrated luminosity of 1000 fb$^{-1}$ will produced more than $3\times 10^4$ and $ 10^5$ $\phi$ for unpolarized and polarized cases, respectively. As long as $Br_{0,X}$ is larger than a few times of $10^{-4}$, $\phi \to X$ may be studied by a unpolarized $\gamma\gamma$ collider.  Depending on the efficiency of identifying the final $\gamma$'s, the $\phi\to \gamma\gamma$ may be confirmed at a $\gamma\gamma$ collider. For the case of a polarized beams, the enhanced production cross section allows one to study properties of $\phi$ to good precisions.

In principle, there should be other possible decay modes other than  
$\phi \to \gamma\gamma$, although at present only $\phi \to \gamma\gamma$ has been observed. 
One can easily measure the relative branching ratios for other decay modes, because~\cite{gammagammacollider}
\begin{eqnarray}
{\sigma^L_{0,X}\over \sigma^L_{0,\gamma\gamma}} = {Br_{0,X}\over Br_{0,\gamma\gamma}} = {\Gamma_{0,X}\over \Gamma_{0,\gamma\gamma}}\;. \label{ratio}
\end{eqnarray}

An immediate interesting measure is to decide whether $g g \to \phi$ is the main production mechanism for $\phi$ at the LHC.
Approximating $\Gamma(\phi \to 2jets) \approx \Gamma(\phi \to gg)$, $\Gamma_{0, gg}$ can be easily determined~\cite{theories3,theories4}.

Assuming $\phi$ is a SM singlet, the interaction inducing $\phi \to \gamma\gamma$ should respect the SM gauge symmetry and
can be parameterized as
\begin{eqnarray}
L = \phi (\tilde a B_{\mu\nu} B^{\mu\nu} + \tilde b W_{\mu\nu} W^{\mu\nu})\;,
\end{eqnarray}
where $B_{\mu\nu}$ and $W_{\mu\nu}$ are the $U(1)_Y$ and $SU(2)_L$ gauge field strengthes, respectively.
If $\phi$ is a pseudoscalar, one replaces $B_{\mu\nu} B^{\mu\nu}$ and $W_{\mu\nu}W^{\mu\nu}$ in the above by $\tilde B_{\mu\nu} B^{\mu\nu}$ and $\tilde W_{\mu\nu}W^{\mu\nu}$, respectively. Here
$\tilde X_{\mu\nu} = (1/2)\epsilon_{\mu\nu\alpha\beta}X^{\alpha\beta}$.

The above Lagrangian will not only induce $\phi \to \gamma\gamma$, but, in general, also $\phi \to \gamma Z, ZZ, W^+W^-$.
None of the later three decays have been observed experimentally. Using eq.(\ref{ratio}), one can study the decay widths for $\phi\to \gamma Z, ZZ, W^+W^-$ at a $\gamma\gamma$ collider~\cite{theories3}. We now look at this in a slightly different way than that carried out in Ref.~\cite{theories3}. For convenience, we normalize $\tilde a (\tilde b) = (4\pi\Gamma/m^3_\phi)^{1/2} a (b)$. We have
\begin{eqnarray}
\Gamma_{0,\gamma\gamma} = \Gamma (a c^2_W + b s^2_W)^2\;,
\end{eqnarray}
where $c_W = \cos\theta_W$ and $s_W = sin \theta_W$.

Taking $\Gamma = \Gamma_{0,\gamma\gamma}$, implies that  $(a c^2_W + b s^2_W)^2 = 1$. One can express $a = (1-b s^2_W)/c^2_W$. We then have
\begin{eqnarray}
R_{\gamma Z/\gamma\gamma} &=& {\sigma^L_{0,\gamma Z}\over \sigma^L_{0,\gamma\gamma}} ={\Gamma_{0,\gamma Z}\over \Gamma_{0,\gamma\gamma}}
= 2\tan^2\theta_W (1-b)^2 (1-m^2_Z/m^2_\phi)^3\;,\nonumber\\
R_{ZZ/\gamma\gamma} &=& {\sigma^L_{0,Z Z}\over \sigma^L_{0,\gamma\gamma}} = {\Gamma_{0,Z Z}\over \Gamma_{0,\gamma\gamma}}
= (\tan^2\theta_W (1-b) + b)^2 (1-4m^2_Z/m^2_\phi)^{3/2}\;,\nonumber\\
R_{W^+W^-/\gamma\gamma} &=& {\sigma^L_{0,W^+W^-}\over \sigma^L_{0,\gamma\gamma}} = {\Gamma_{0,W^+W^-}\over \Gamma_{0,\gamma\gamma}}
= 2b^2(1-4m^2_W/m^2_\phi)^{3/2}\;.
\end{eqnarray}

\begin{figure}
\includegraphics[width=7cm]{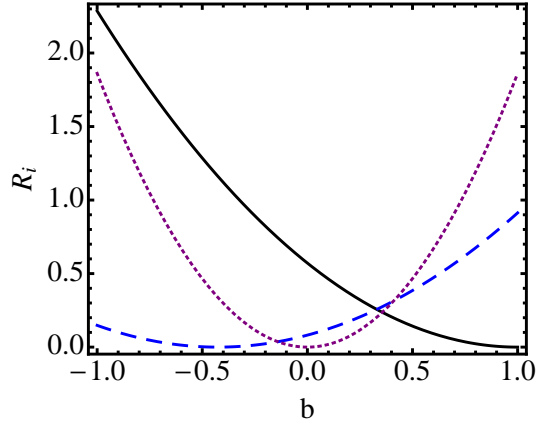}
\caption{$R_i$ as functions of $b$. The solid, dashed and dotted curves are for $R_{\gamma Z/\gamma\gamma}$, 
$R_{ZZ/\gamma\gamma}$ and $R_{W^+W^-/\gamma\gamma}$, respectively.}\label{fig3}
\end{figure}

In Figure \ref{fig3}, we show $R_i$ as functions of $b$. Note that one cannot simultaneously make all $R_i$ to be zero.  
At least two of them will show up at some level. The worst scenario is that b is about 0.3 where all three $R_i$ are below 0.25. 
This may be the reason that these decay modes have not been observed at the LHC run II.
If this is indeed the case, these decay modes may be difficult to be studied at a unpolarized $\gamma\gamma$ collider for $\Gamma_{0,\gamma\gamma}$ taking its present lower bound. But for the polarized case, it is still possible. Once one of the $R_i$ is measured, the other two will be predicted. 
One should be aware that at the LHC, the production mechanism may not be due to $gg\to \phi$. Photo-production may also be possible~\cite{photo-production, photo-production1}. In that case $\Gamma_{0,\gamma\gamma}$ can be much larger than 1 MeV. A $\gamma\gamma$ collider with unpolarized photon beams can also study the decay modes to good precision.  We conclude that it is possible to verify the existence of $\phi$ at a $\gamma\gamma$ at a CM frame energy of 1 TeV with an integrated luminosity of 1000 fb$^{-1}$.
\\

\noindent 
{\bf Angular distribution of $\gamma$ and spin of $\phi$}

Since the LHC has observed $\phi \to \gamma\gamma$, $\phi$ cannot be a spin-1 particle due to Landau-Yang theorem~~\cite{landau}. The two possible low spin states $0$ and $2$ are not ruled out. The consequences should be studied~\cite{spin-2}. A $\gamma\gamma$ collider can provide detailed information about the spin of $\phi$ by studying the angular distribution of the final photons.

For a scalar or a pseudoscalar $\phi$, the angular distribution for one of the final photon respecting to the incoming $\gamma$ beams shown in Fig.\ref{fig4} is isotropic in the $\gamma\gamma$ CM frame~~\cite{gammagammacollider,Choi}
\begin{eqnarray}
{1\over \sigma(s)_{0, \gamma\gamma}}{d\sigma(s)_{0, \gamma\gamma}\over d \cos\theta} = 1\;. \label{dis}
\end{eqnarray}

In the $e^+e^-$ CM frame (laboratory frame), collision of the two photons  is not in the $\gamma\gamma$ CM frame and therefore
the distribution of the photons is not the same as that predicted by eq. (\ref{dis}). In the laboratory frame, depending on the values of $x_{1}$ and $x_{2}$, the two photons may have different energy. The $\phi$ produced will be boosted to the direction of the photon with a larger $x_i$. The angle $\theta$ when seeing from laboratory frame will be changed to $\theta_L$. The relation between $\theta$ and $\theta_L$ can be written as the following
\begin{eqnarray}
\cos \theta = \frac{\cos \theta_{L}+\beta}{\beta\cos \theta_{L}+1}\;,\;\;\frac{d\cos \theta}{d\cos \theta_{L}} = \frac{1-\beta ^{2}}{(\beta\cos \theta_{L}+1)^{2}}\;,\;\;\beta = \frac{x_{1}-x_{2}}{x_{1}+x_{2}}\;. \label{boost}
\end{eqnarray}

The laboratory frame angular distribution $A(0,\theta_L)$ of $\theta_{L}$ for a spin-0 scalar can be studied by the following convoluted distribution,
\begin{eqnarray}
A(0, \theta_L) = {1\over \sigma(s)_{0, \gamma\gamma}^L} \int^{x_{max}}_{x_{min}} d x_1 \int^{x_{max}}_{x_{min}}dx_2  f(x_1)f(x_2) {d\sigma(x_1 x_2 s)_{0, \gamma\gamma}\over d \cos\theta_{L}}\;.
\end{eqnarray}

One can also defined a similar quantity for the case of $\phi$ being a particle with a different spin, spin-J, $A(J, \theta_L)$. We find that this quantity can give information about the spin of $\phi$.
To see how this works, we take an example of a spin-2  tensor coupled to $\gamma\gamma$, in a similar fashion as a scalar couples to two gravitons, to study $A(2,\theta_L)$ and compared with $A(0,\theta_L)$ for a spin-0 scalar. In this case the matrix element for $\phi \to \gamma\gamma$ can be written as~~\cite{Han:1998sg}
\begin{eqnarray}
&&M(2, \gamma\gamma) = \frac{-\kappa}{2}[(k_{1}\cdot k_{2})C_{\mu\nu,\varrho\sigma}+D_{\mu\nu,\varrho\sigma}(k_{1},k_{2})]\varepsilon ^{\rho *}(k_{1})\varepsilon ^{\sigma *}(k_{2})\epsilon^{\mu\nu }\;,\nonumber\\
&&C_{\mu\nu,\varrho\sigma}=\eta_{\mu\varrho}\eta_{\nu\sigma}+\eta_{\mu\sigma}\eta_{\nu\rho}
-\eta_{\mu\nu}\eta_{\rho\sigma}\;,\\
&&D_{\mu\nu,\varrho\sigma}(k_{1},k_{2}) = \eta_{\mu\nu}k_{1\sigma}k_{2\rho}-[\eta_{\mu\sigma}
k_{1\nu}k_{2\rho}+\eta_{\mu\rho}k_{1\sigma}k_{2\nu}-\eta_{\rho\sigma}k_{1\mu}k_{2\nu}+
(\mu\leftrightarrow\nu)]\;.\nonumber
\end{eqnarray}

In the $\gamma\gamma$ CM frame, we have~\cite{gammagammacollider}
\begin{eqnarray}
{1\over \sigma(s)_{2,\gamma\gamma}}{d\sigma(s)_{2,\gamma\gamma}\over d\cos\theta} ={5\over 16}(\cos^4\theta+ 6\cos^2 \theta + 1)\;.
\end{eqnarray}

Note that in the $\gamma\gamma$ CM frame, the final $\gamma$ has a non-trivial angular distribution for spin-2 tensor. This also shows up in the laboratory frame. We find that if comparing $A(0, \theta_L)$ and $A(2,\theta_L)$, one can distinguish different cases even without knowing  $\Gamma_{i, \gamma\gamma}$ and $B_{i, \gamma\gamma}$  separately.

In Fig.\ref{fig5}, we plot $A(0, \theta_L)$ and $A(2,\theta_L)$ for several different $\sqrt{s}$. 
We see that at
$\sqrt{s} = 1$ TeV, the differences for spin-0 and spin-2 laboratory frame angular distribution are substantial. This can be used to distinguish whether $\phi$ is a spin-0 or a spin-2 state. For spin-0 case, the distribution is almost flat despite of the boost effect shown in eq.(\ref{boost}) in the laboratory frame. This is because that at 1~TeV, the particle produced has a small kinetic energy and it is almost at rest. However, the boost effects show up with higher energies which can be clearly seen in Figures 5.b and 5.c where the laboratory frame energies are 1.5~TeV and 2~TeV, respectively. Also note that at 1~TeV laboratory frame energy, there is almost no difference between the polarized and unpolarized cases despite the distributions for these two cases are different and therefore the boost effects should be different. Again this is because that at 1~TeV laboratory energy, the boost effects are small. At higher energies, the effects become visible as can be seen from Figures 5.b and 5.c. 

For the purpose of studying the 750~GeV resonant state $\phi$, if energy is higher than 1~TeV, the production cross section drops and therefore event rate becomes smaller. So in practice, one prefers energy not much larger than 1~TeV. Also the angular differences between the spin-0 and spin-2 cases become smaller when energies go higher. Energy higher than 1~TeV is, again, not favored. We now take $\sqrt{s}=1$~TeV to estimate theoretical statistic error in angular distribution measurements.  We will assume the total integrated luminosity to be 1000 fb$^{-1}$ as our input. 

To obtain event numbers, one needs to know $\Gamma_{J,\gamma\gamma}^2/\Gamma_{J,total}$. We use 
$\Gamma_{J, gg}\Gamma_{J,\gamma\gamma} /\Gamma_{J,total} \sim 1$~MeV as an input for our estimate. Theoretical estimate of $\delta = \Gamma_{J,\gamma}/\Gamma_{J, gg}$ has a large range. For example, in scenarios of a warped extra dimension containing bulk SM fields, $\delta$ is estimated to be of order~\cite{hewett} 0.1 for $\phi$ with spin-2, and in other models, it ranges from  $10^{-2}$ to $10^{-3}$~\cite{theories4}. We will take a middle value $\delta = 10^{-2}$ as an example for estimate. In this case $\Gamma_{J,\gamma\gamma}^2/\Gamma_{J,total} \sim 10^{-2}$~MeV. With the total integrated luminosity to be 1000 fb$^{-1}$, with information in Figure {\ref{fig2} one would obtain an event number $N \sim 290$ for unpolarized case, and $N \sim 1070$ for polarized case for $\sqrt{s}=1$ TeV. 
In Figure \ref{fig6} we show histogram plots for $N_i/N$ as a function of $\cos\theta_L$ with $N_i$ being the event number in a bin for an interval of 0.2 for $\cos\theta_L$. 
 
From Figure \ref{fig6}, we see that the separation at $\cos\theta_L$ close to 0 or $\pm 1$ for unpolarized case in some bins the significance can be about 2$\sigma$, but for polarized case separation can be much more significant as can be seen in Figure \ref{fig6}.
Note that in $N_i/N$ the factor product of luminosity times $\Gamma_{J,\gamma\gamma}^2/\Gamma_{J,total}$ is cancelled out, it does not depend on $N$. The error in $N_i/N$ scales as $1/\sqrt{N}$. Therefore with higher luminosity or larger $\delta$, the error bars will shrink and make the distinction of spin-0 and spin-2 cases even more obvious. In the worst scenario case where $\Gamma_{\gamma\gamma} = 1$~MeV and $\Gamma_{total} = \Gamma_{gg} = 45$~GeV, with integrated luminosity of 1000fb$^{-1}$, even for polarized case the event number is only a few, there is not enough statistics for deciding whether $\phi$ has spin-0 or spin-2. A much larger luminosity is needed.

\noindent
{\bf Summary}

To summarize, we have studied how a $\gamma\gamma$ collider can help to provide some of the most important information about the $\phi$ resonant state hinted by LHC run II data. We have shown that a $\gamma\gamma$ collider constructed by using the laser backscattering
technique on the electron and positron beams in an $e^+e^-$ collider can verify whether $\phi$ indeed exists, and probe some of important 
properties of it.  The optimal $\sqrt{s}$ is slightly below 1 TeV. With an integrated luminosity of 1000 fb$^{-1}$, different models can be tested by studying  $\phi \to \gamma Z$, $\phi \to Z Z$ and $\phi \to W^+W^-$.  Studying angular distribution of the $\gamma \gamma$ through on-shell production of $\phi$ and its subsequent decays into a $\gamma \gamma$ pair can provide useful information whether the $\phi$ is  a spin-0 or a spin-2 state.

\begin{figure}
\includegraphics[width=6cm]{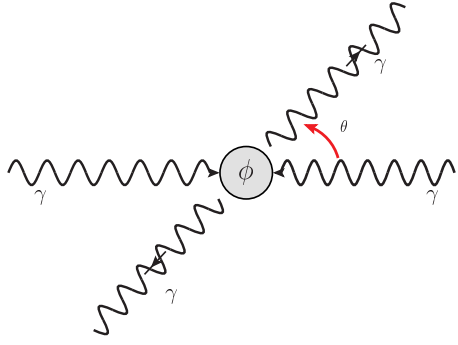}
\caption{The angle $\theta$ of a final photon in $\gamma\gamma \to \phi \to \gamma\gamma$.}\label{fig4}
\end{figure}

\begin{figure}
\includegraphics[width=5cm]{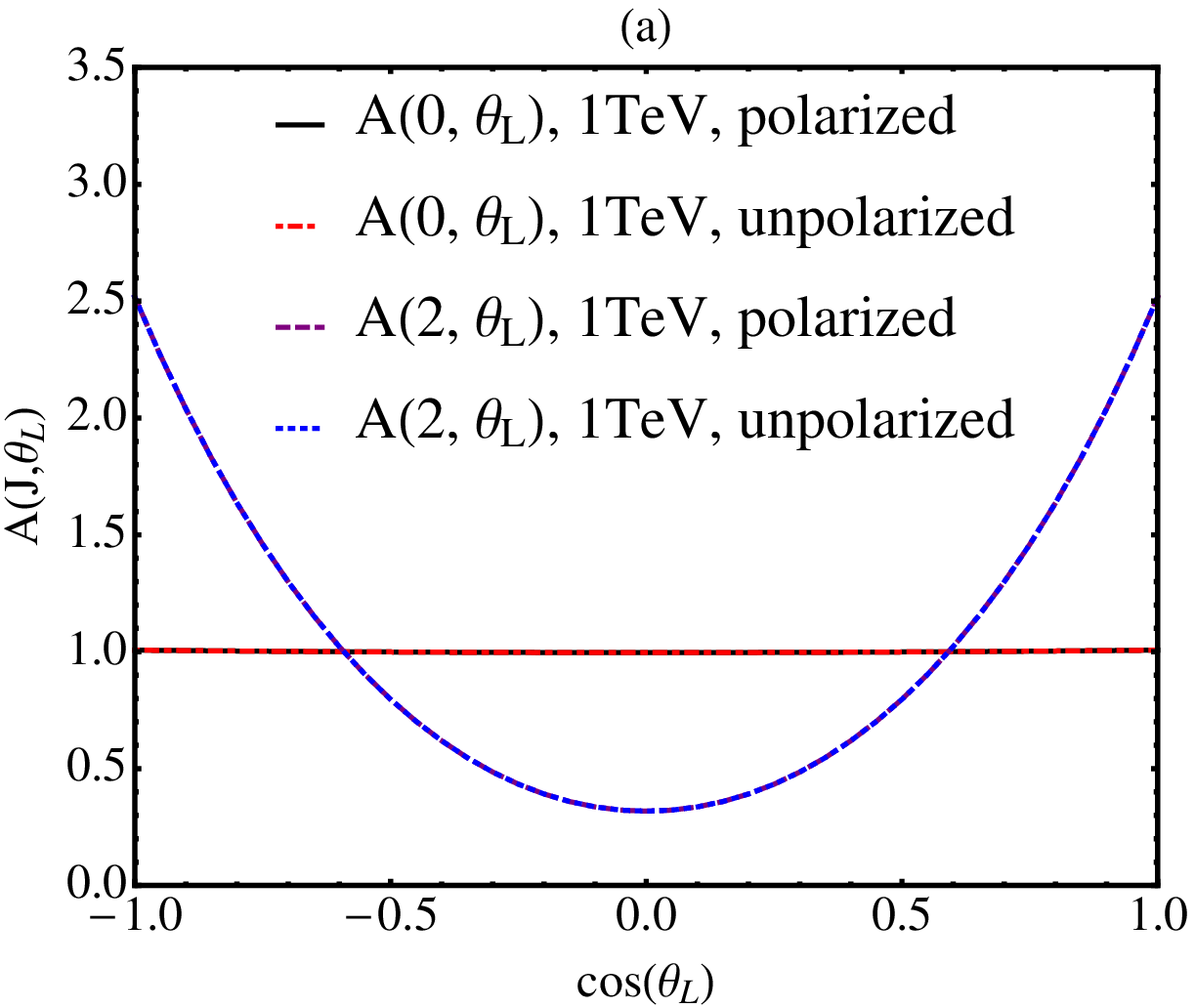}
\includegraphics[width=5cm]{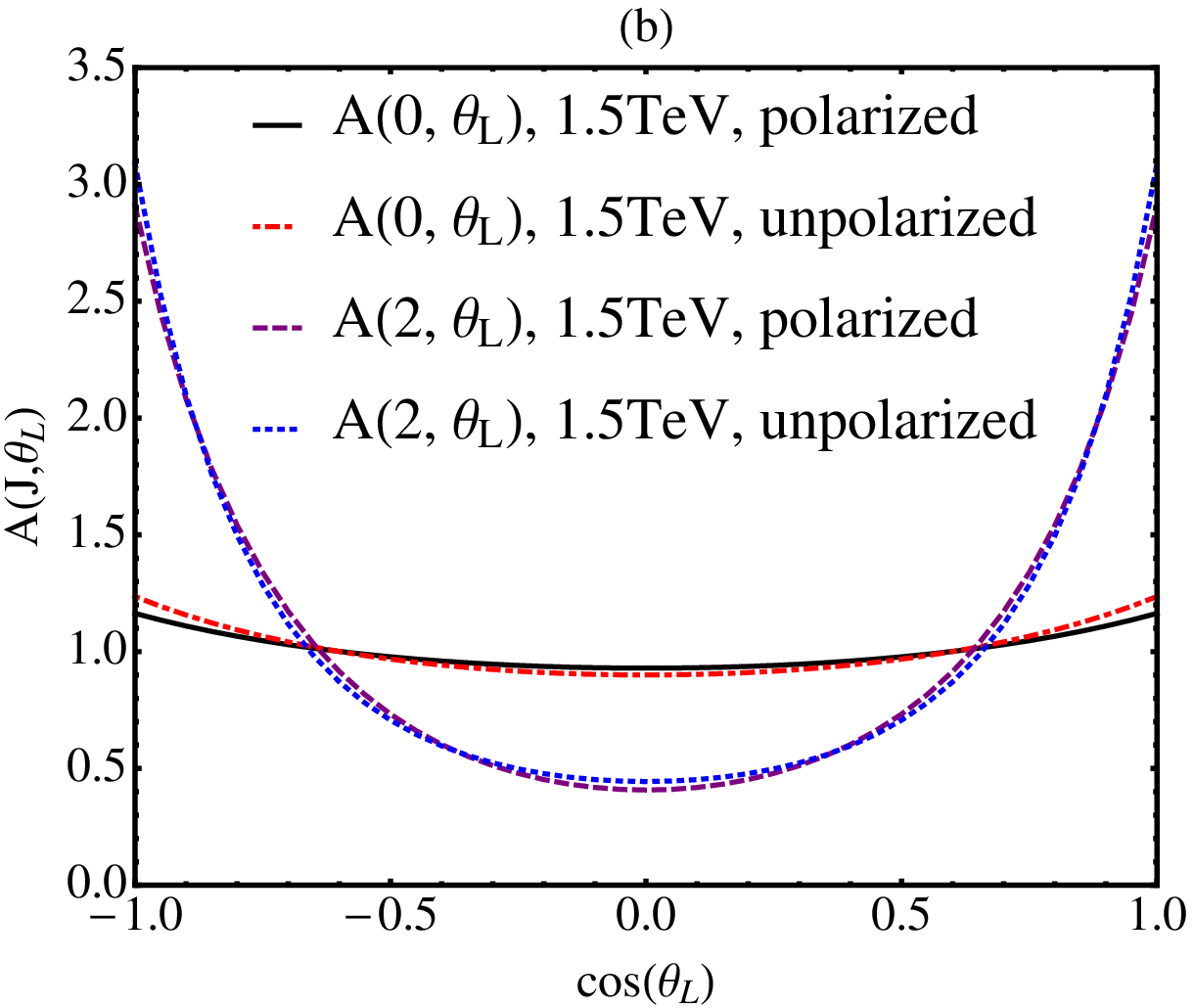}
\includegraphics[width=5cm]{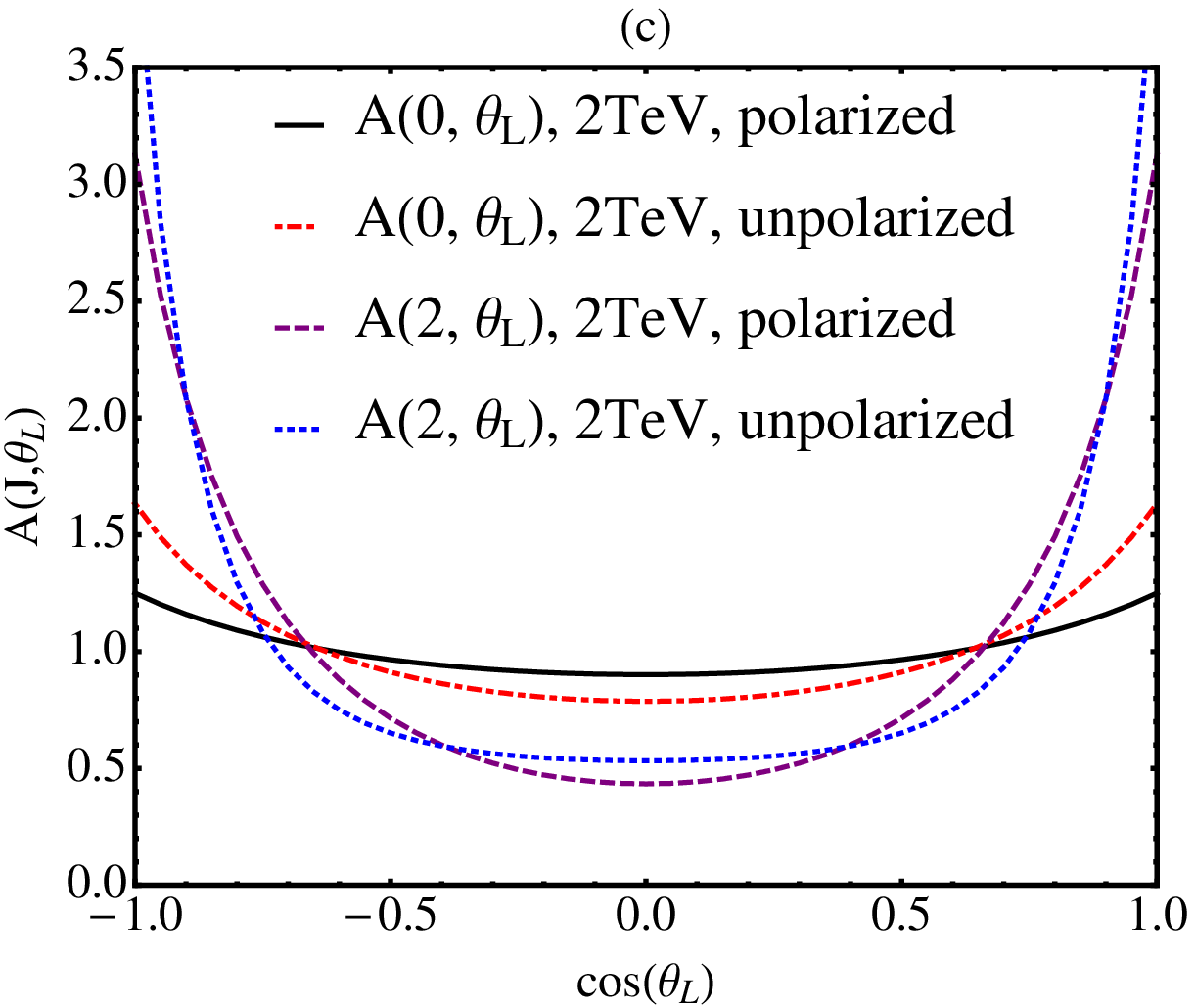}
\caption{Angular distribution $A(J=0,2, \gamma\gamma)$ in $\gamma\gamma \to \phi \to \gamma \gamma$ with different 
energies $\sqrt{s}$.}\label{fig5} 
\end{figure}

\begin{figure}
\includegraphics[width=7cm]{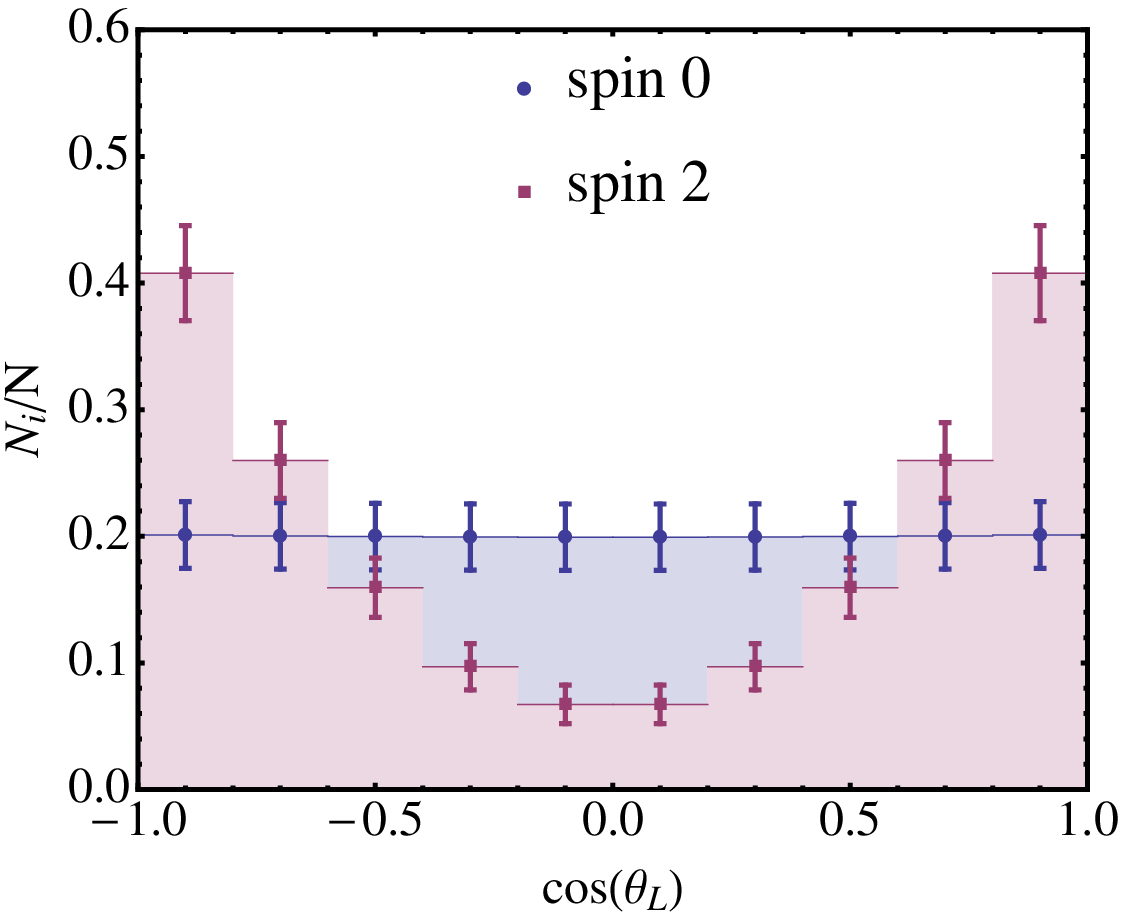}
\includegraphics[width=7cm]{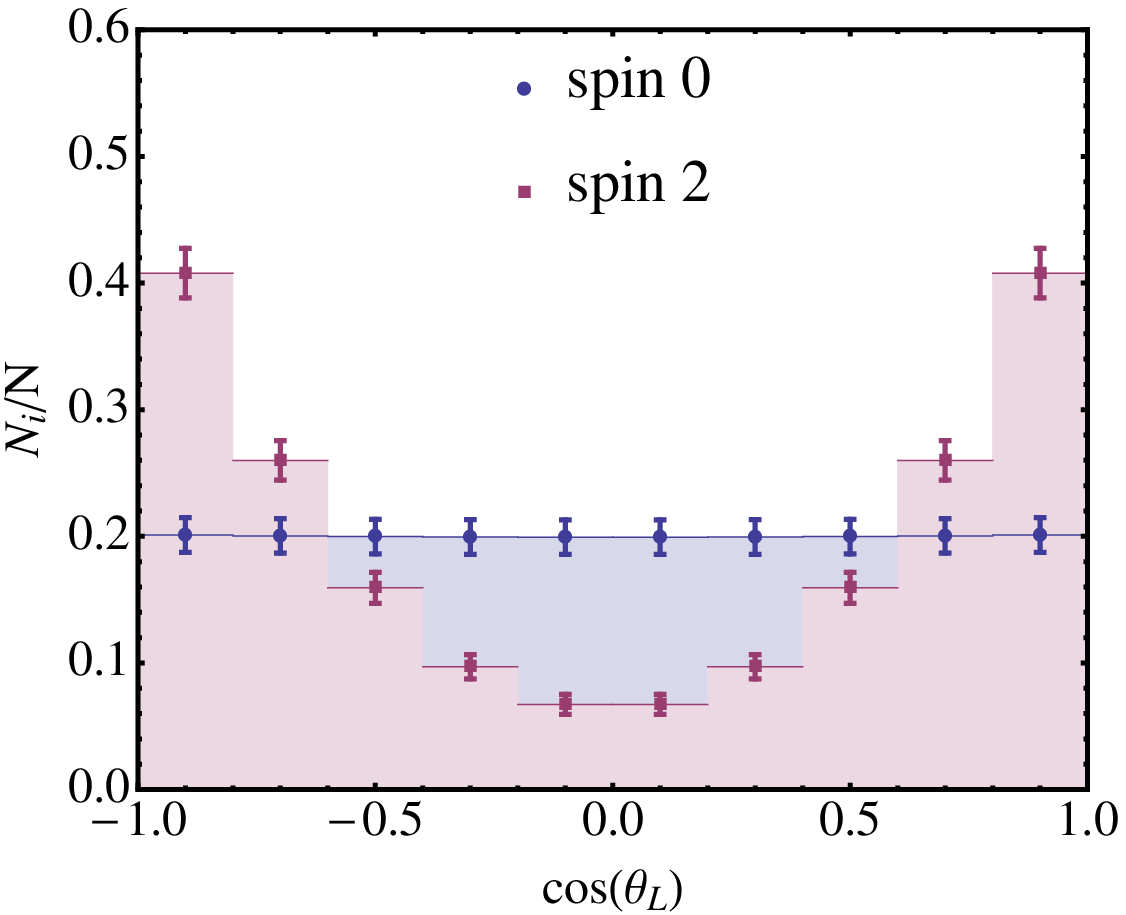}
\caption{Histogram plots for $N_i/N$ distributions for spin-0 and spin-2 cases with a fixed total event number $N$. Left and right panel are for $N= 290$ and $N=1070$, respectively.}\label{fig6} 
\end{figure}

\acknowledgments \vspace*{-1ex}
X-G He was supported in part by MOE Academic Excellent Program (Grant No.~102R891505) and MOST of ROC (Grant No.~MOST104-2112-M-002-015-MY3), and in part by NSFC (Grant Nos.~11175115 and 11575111) and Shanghai Science and Technology Commission (Grant No.~11DZ2260700) of PRC. X.~G.~H. thanks Korea Institute for Advanced Study (KIAS) for their hospitality and partial support while this work was completed.

\end{document}